\documentclass[twocolumn,showpacs,aps,prl,superscriptaddress,floatfix]{revtex4}

\usepackage{graphicx}
\usepackage{dcolumn}
\usepackage{amsmath}
\usepackage{epsfig}

\input{babarsym.tex}

\def\stwobg   {\ensuremath{\sin(2\beta+\gamma)}\xspace}
\def\stwobgd   {\ensuremath{\sin(2\beta+\gamma\pm\delta)}\xspace}
\def\KKstarz  {\ensuremath{K^{(*)0}}\xspace}
\def\KKstarzb {\ensuremath{\Kbar^{(*)0}}\xspace}

\def\DDstarz {\ensuremath{D^{(*)0}}\xspace}
\def\DDstarzb{\ensuremath{\Dbar^{(*)0}}\xspace}

\providecommand{\KPi}{\mbox{\ensuremath{K^-\pi^+}}}
\providecommand{\KPiPiPi}{\mbox{\ensuremath{K^-\pi^+\pi^-\pi^+}}}
\providecommand{\KPiPiz}{\mbox{\ensuremath{K^-\pi^+\pi^0}}}

\providecommand{\BR}{\mbox{\ensuremath{\mathcal B}}}

\def\fisher {\ensuremath{\mathcal F}\xspace}

\def\rbk {\ensuremath{\tilde{r}_B}}
\def\Kbar    {\kern 0.2em\overline{\kern -0.2em K}{}\xspace}

\def\Btilde {\ensuremath{\tilde{B}}}
\def\Bztilde {\ensuremath{\Btilde^{0}}}
\def\Ktilde {\ensuremath{\tilde{K}}}
\def\KKstarztilde  {\ensuremath{\Ktilde^{(*)0}}\xspace}
\def\Kztilde  {\ensuremath{\Ktilde^{0}}\xspace}

\newcommand{\BABARPubYear}    {06}
\newcommand{\BABARPubNumber}  {002}

\newcommand{\SLACPubNumber}   {11806}

\def\figurebox#1#2#3{%
    \def\arg{#3}%
    \ifx\arg\empty
    {\hfill\vbox{\hsize#2\hrule\hbox to #2{\vrule\hfill\vbox to #1{\hsize#2\vfill}\vrule}\hrule}\hfill}%
    \else
    {\hfill\epsfbox{#3}\hfill}%
    \fi}

\begin{document}

\preprint{\babar-PUB-\BABARPubYear/\BABARPubNumber}
\preprint{SLAC-PUB-\SLACPubNumber}

\begin{flushleft}
\babar-PUB-\BABARPubYear/\BABARPubNumber\\
SLAC-PUB-\SLACPubNumber\\
%hep-ex/\LANLNumber\\
\end{flushleft}

\title{
{\large \bf Measurement of $\Bzb\ra \DDstarz \KKstarzb$ Branching Fractions} }

%% author list as of 02-Jan-2006 (616 authors)
%
\author{B.~Aubert}
\author{R.~Barate}
\author{D.~Boutigny}
\author{F.~Couderc}
\author{Y.~Karyotakis}
\author{J.~P.~Lees}
\author{V.~Poireau}
\author{V.~Tisserand}
\author{A.~Zghiche}
\affiliation{Laboratoire de Physique des Particules, F-74941 Annecy-le-Vieux, France }
\author{E.~Grauges}
\affiliation{Universitat de Barcelona, Facultat de Fisica Dept. ECM, E-08028 Barcelona, Spain }
% ...
\collaboration{The \babar\ Collaboration}
\noaffiliation

\date{\today}% It is always \today, today, but you may specify any date
             % with \date.

%%%%%%%%%%%%%%%%%%%%%%%%%%%%%%%
% all numbers are defined here
%
\def\lumi {205.5}
\def\BBcount {226}
\def\offlumi {11.9}
%
% D0 K0
\def\dzkzBrVal  {5.3}
\def\dzkzBrStat {0.7}
\def\dzkzBrSyst {0.3}
%
% D*0 K0bar
\def\dstarzkzBrVal  {3.6}
\def\dstarzkzBrStat {1.2}
\def\dstarzkzBrSyst {0.3}
%
% D0bar K*0 (Vcb)
\def\dzbkstarzBrVal  {4.0}
\def\dzbkstarzBrStat {0.7}
\def\dzbkstarzBrSyst {0.3}
%
% D0 K*0 (Vub)
\def\dzkstarzBrVal  {0.0}
\def\dzkstarzBrStat {0.5}
\def\dzkstarzBrSyst {0.3}
\def\dzkstarzLim    {1.1}
\def\brscale{\ensuremath{\times10^{-5}}}
%%%%%%%%%%%%%%%%%%%%%%%%%%%%%%%

\begin{abstract}
We present a study of the  decays $\Bzb\ra\DDstarz\KKstarzb$ using
a sample of \BBcount\ million $\Y4S\ra B\Bbar$ decays collected
with the \babar\ detector at the \pep2\ asymmetric-energy \epem\
collider at SLAC. We report evidence for the decay of \Bz\ and
\Bzb\  mesons to the $\Dstarz\KS$ final state with an average
branching fraction
$\BR(\Bztilde\ra\Dstarz\Kztilde)\equiv
( \BR(\Bzb\ra\Dstarz\Kzb) + \BR(\Bz\ra\Dstarz\Kz) )/2
=(\dstarzkzBrVal\pm\dstarzkzBrStat\pm\dstarzkzBrSyst)\brscale$.
Similarly, we measure
$\BR(\Bztilde\ra\Dz\Kztilde)\equiv
( \BR(\Bzb\ra\Dz\Kzb) + \BR(\Bz\ra\Dz\Kz) )/2
=(\dzkzBrVal\pm\dzkzBrStat\pm\dzkzBrSyst)\brscale$ 
for the $\Dz\KS$ final state. We measure $\BR(\Bzb\ra\Dz\Kstarzb)
=(\dzbkstarzBrVal\pm\dzbkstarzBrStat\pm\dzbkstarzBrSyst)\brscale$
and set a 90\% confidence level upper limit $\BR(\Bzb\ra\Dzb\Kstarzb) <
\dzkstarzLim\brscale$.
We determine the upper limit for the decay amplitude ratio 
$|{\mathcal A}(\Bzb\ra\Dzb\Kstarzb)/{\mathcal A}(\Bzb\ra\Dz\Kstarzb)|$
to be less than 0.4 at the 90\% confidence level.
\end{abstract}

\pacs{13.25.Hw, 12.15.Hh, 11.30.Er}% PACS, the Physics and Astronomy
                                   % Classification Scheme.

%%%%%%%%%%%%%%%%%%%%%%%%%%%%%%%%%%%%%%%%%%%%%%%%%%%%%%%%%%%%%%%%%%%%%%%%%%%%%%%
%  please do not use more than 80 columns which                              80
%%%%%%%%%%%%%%%%%%%%%%%%%%%%%%%%%%%%%%%%%%%%%%%%%%%%%%%%%%%%%%%%%%%%%%%%%%%%%%%
\maketitle

\begin{figure}[htb]
\begin{center}
\epsfig{file=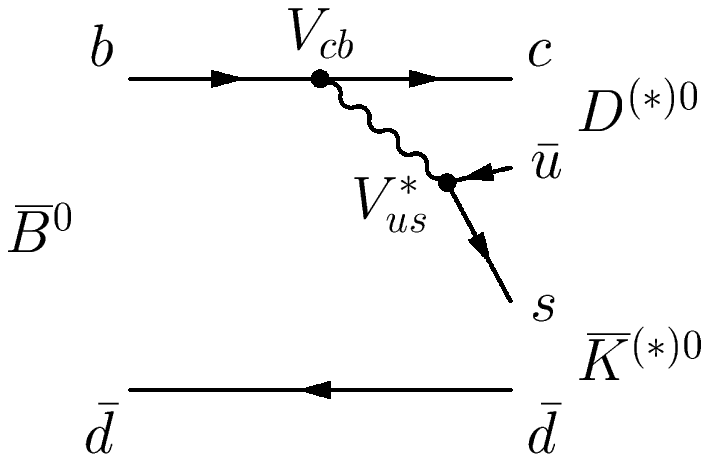,width=0.49\linewidth}
\epsfig{file=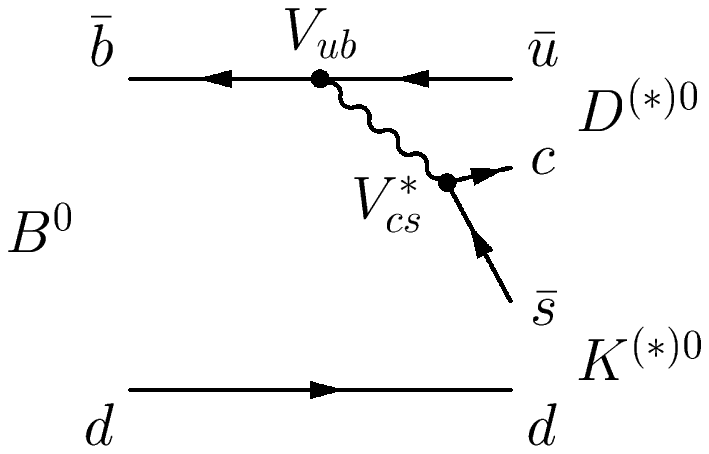,width=0.49\linewidth}
\end{center}
\caption{
The decay diagrams for the $b\ra c$ transition $\Bzb\ra\DDstarz\Kzb$
and the $\bbar \ra \ubar$ transition $\Bz\ra\DDstarz\Kz$.
}
\label{fig:feyn}
\end{figure}

With the discovery of \CP\ violation in the decays of neutral $B$
mesons~\cite{s2b-discovery} and the precise
measurement~\cite{sin2beta} of the angle $\beta$ of the
Cabibbo-Kobayashi-Maskawa~(CKM) Unitarity Triangle~\cite{CKM}, the
experimental focus has shifted
towards over-constraining the unitarity triangle through precise 
measurements of \Vub\ and the angles $\alpha$ and $\gamma$.
The angle $\gamma$ is ${\rm arg}(-V_{ub}^{*}V_{ud}/V_{cb}^{*}V_{cd})$ and $V_{ij}$ 
are CKM matrix elements.
Several methods have been suggested and explored to
measure $\gamma$ with small uncertainties~\cite{gamma-status}, but they all require large
samples of $B$ mesons not yet available.
The decay modes $\Bzb\ra\DDstarz\Kzb$ offer a new approach for the
determination of \stwobg\ from the measurement of time-dependent
\CP\ asymmetries in these decays~\cite{theory-DK}.
The \CP\ asymmetry appears as a  result of the interference
between two diagrams leading to the same final state
$\DDstarz\KS$~(Figure~\ref{fig:feyn}).
A $\Bzb$ meson can either decay via a $b\ra c$ quark transition to
the $\DDstarz\Kzb$ ($\Kzb\to\KS$) final state, or oscillate into a
$\Bz$ which then decays via a $\bar b\ra \bar u$ transition to the
$\DDstarz\Kz$ ($\Kz\to\KS$) final state~\cite{charge-conj}.
The $\Bzb\Bz$ oscillation provides the weak phase $2\beta$ and
the relative  weak phase between the two decay diagrams is $\gamma$.

The sensitivity of this method~\cite{theory-DK} depends on the rates 
for these decays and the ratio of the decay amplitudes.
The branching fractions ${\mathcal B}(\Bzb\ra\DDstarz\KKstarzb)$ 
can be estimated from the measured color-suppressed
decays $\Bzb\ra\DDstarz\piz$~\cite{color-suppressed} to be
approximately 
${\mathcal B}(\Bzb\to \DDstarz\KKstarzb)\approx \sin^2\theta_{c} \ 
{\mathcal B}(\Bzb\to \DDstarz\pi^0) \sim {\cal O}(10^{-5})$,
where $\theta_{c}$ is the Cabibbo angle and $\sin\theta_{c}=0.22$.
The Belle Collaboration has observed 
the $\Bzb\ra\Dz\KKstarzb$ decays with branching fractions consistent with this
naive expectation~\cite{belle-d0ks-prl}. The time-dependent \CP\ asymmetries in
$\Bzb\ra\DDstarz\Kzb$
decays are proportional to
${r^{(*)}_B}\cdot\stwobgd/(1+{r^{(*)}_B}^2)$,
 where 
$r^{(*)}_B\equiv|{\mathcal A}(\Bzb\ra\DDstarzb\Kzb)/{\mathcal A}(\Bzb\ra\DDstarz\Kzb)|$ and
$\delta$ is a relative strong phase which depends on the specific final state.
Higher values of $r^{(*)}_B$ lead to 
larger interference between the  $b\ra c$ and  $b\ra u$ processes 
and thus increased sensitivity to 
the angle $\gamma$. In the Standard Model 
$r^{(*)}_B= f \cdot  |V_{ub} V_{cs}^{*}| / |V_{cb} V_{us}^{*}|$, 
where the factor $f$
accounts for the difference in the strong interaction dynamics 
between 
the  $b\ra c$ and  $b\ra u$ processes. There are no theoretical calculations
or experimental constraints on $f$.

In $\Bzb\ra\DDstarz\Kzb$ ($\Kzb\to\KS$) decays the
strangeness content of the \Kzb\ is hidden and  one cannot
distinguish between $\Bzb\ra\DDstarz\Kzb$ and $\Bz\ra\DDstarz\Kz$.
Therefore a direct determination of $r^{(*)}_B$ from the measured rates is not feasible. 
In the remainder of this paper we refer to these decays as
$\Bztilde\ra\DDstarz\Kztilde$. 
Insight into the $B$ decay dynamics affecting $r^{(*)}_B$ can be 
gained by measuring a similar amplitude ratio 
$\rbk\equiv|{\mathcal A}(\Bzb\ra\Dzb\Kstarzb)/{\mathcal A}(\Bzb\ra\Dz\Kstarzb)|$ 
using the self-tagging decay
$\Kstarzb\ra\Km\pip$. The $\Bzb\ra\Dz\Kstarzb$ and
$\Bzb\ra\Dzb\Kstarzb$ decays are distinguished by the
correlation between the charges of the kaons produced in the
decays of the neutral $D$ and the \Kstarzb. In the former decay
the two kaons in the final state must have the same charge, while in the latter they
are oppositely charged.
This charge correlation in the final state is diluted by the presence of the 
doubly-Cabibbo-suppressed decays
$\Dz\ra\Kp\pim,\Kp\pim\piz$, and $\Kp\pim\pip\pim$.
The ratio \rbk\ is related to the experimental observables ${\mathcal R}_i$ defined as
\begin{eqnarray}
{\mathcal R}_i & = & \frac{\Gamma(\Bzb\ra(\Kp X_i^{-})_{D}\Kstarzb)}
                          {\Gamma(\Bzb\ra(\Km X_i^{+})_{D}\Kstarzb)}\nonumber \\
             & = & \rbk^{2} + r_{D_i}^{2} + 2\rbk r_{D_i}\cos(\gamma+\delta_i)\label{R_ADS} ,
\end{eqnarray}
where
\begin{eqnarray}
X_i^{\pm} & = & \pi^{\pm}, \pi^{\pm}\piz, \pi^{\pm}\pim\pip,\\
r_{D_i} & = & \frac{|{\mathcal A}(\Dz\ra\Kp X_i^{-}  )|}
                 {|{\mathcal A}(\Dz\ra\Km  X_i^{+} )|},\\
 \delta_i & = & \delta_{B} + \delta_{D_i}, 
\end{eqnarray}
and $\delta_{B}$ and $\delta_{D_i}$ are strong phase differences between the two $B$ and
$D_i$ decay amplitudes, respectively. The values of $r_{D_i}$ have been measured 
to be $r_{D\ra K\pi} = 0.060\pm0.002$,  $r_{D\ra K\pi\piz} = 0.066\pm 0.010$, and
$r_{D\ra K\pi\pi\pi} = 0.065\pm 0.010$~\cite{PDG}.

We present herein
measurements of the  branching fractions
${\mathcal B}(\Bztilde\ra\Dz\Kztilde)$
and ${\mathcal B}(\Bzb\ra\Dz\Kstarzb)$,
evidence for the decay 
$\Bztilde\ra\Dstarz\Kztilde$,
a $90\%$ confidence level (C.L.) upper limit for the branching fraction of the
$b\ra u$ transition $\Bzb\ra\Dzb\Kstarzb$, and
a limit for the ratio \rbk.

These results
 are based on a sample of \BBcount\ million
$\Y4S \ra \BB$ decays collected with the \babar\
detector between 1999 and 2004 at the \pep2\ asymmetric-energy
\epem\ collider operating at the \Y4S\ resonance.
The properties of the continuum $\epem\ra q\bar q \ (q=u,d,s,c)$ background
events are studied with  a data sample of \offlumi~\invfb\
recorded at an energy 40~\mev\ below the \Y4S\ resonance.
The \babar\ detector has been described in detail elsewhere~\cite{babar-detector-nim}.
Detector components relevant
for this analysis are summarized here. Trajectories of charged
particles are measured in a spectrometer consisting of a
five-layer silicon vertex tracker (SVT) and a 40-layer drift
chamber (DCH) operating in a 1.5~T axial magnetic field.  Charged
particles are identified as pions or kaons using information from
a detector of internally reflected Cherenkov light, as well as
measurements of energy loss from ionization ($dE/dx$) in the SVT
and the DCH. Photons are detected using an electromagnetic
calorimeter composed of 6580 thallium-doped CsI crystals. We
use a Monte Carlo simulation of the \babar\ detector based on
GEANT4~\cite{Geant4} 
to validate the analysis procedure and to study 
 the backgrounds.
Simulated events are generated with the EvtGen~\cite{evtgen} event generator.

We reconstruct the decays
$\Bzb\ra\Dz\Kzb$, $\Dstarz\Kzb$,
$\Dz\Kstarzb$, and $\Dzb\Kstarzb$ in the decay chains:
$\Dstarz\ra\Dz\piz$; $\Dz\ra\KPi$, \KPiPiz, and \KPiPiPi;
$\Kzb\ra\KS\ra\pipi$; $\Kstarzb\ra\Km\pip$; and
$\piz\ra\gamma\gamma$. 
For each $B$ decay channel the optimal selection criteria are determined
by maximizing the ratio $N_S/\sqrt{N_S+N_B}$, where $N_S$ and $N_B$ are,
respectively, the expected signal and background yields estimated from
samples of simulated events. 
A large sample of the more abundant $\Bp\ra\Dzb\pip$ decays, in
which the $\Dzb$ decays to the  $K^+\pi^-,\ K^+\pi^-\pi^0,$ or
$K^+\pi^-\pi^+\pi^-$ final states, is used as a calibration sample
to measure efficiencies and experimental resolutions for the selection
variables.

Well reconstructed charged tracks are used to  reconstruct \Dz\ and \Kstarz\ candidates.
The \Kpm\ candidates must satisfy a set of kaon identification criteria.

These identification criteria have an average efficiency of about
90\%, while the probability of a pion to be misidentified as a kaon
varies between a few percent and 15\%.
Photons are reconstructed from energy deposition clusters in the
electromagnetic calorimeter consistent with photon showers, and
are required to
have an energy greater than 30~\mev . We select \piz\ candidates from 
pairs of photon candidates by requiring  their invariant mass to be 
in the interval $115~\mevcc<m(\gamma\gamma)<150$~\mevcc.

The \KS\ candidates are selected from pairs of oppositely charged
tracks with invariant mass within 7~\mevcc~($\sim2\sigma$) of the
nominal \KS\ mass. The displacement
of the
\KS\ decay vertex from the interaction point, in the plane
perpendicular to the beam axis, divided by its estimated uncertainty
must be greater than 2.
The \Kstarz\ candidates are selected from pairs of oppositely charged
\Kp\ and \pim\ tracks, with invariant mass within 50~\mevcc\
of the nominal \Kstarz\ mass.
The polarization of the \Kstarz\ in the \Bz\ decay is used to reject 
backgrounds by requiring $|\cos{\theta_{h}}|>0.4$, where the helicity
angle $\theta_h$ is defined as the angle between the direction of the
\Kstarz\ in the \Bz\ meson  rest frame and the direction of its daughter 
\Kp\ in the \Kstarz\ rest frame. For $\Bzb\ra\Dz\Kstarzb$ and
$\Bzb\ra\Dzb\Kstarzb$ signal candidates, $\theta_{h}$ follows a 
$\cos^2\theta_{h}$ distribution, while the combinatorial background is  
distributed uniformly. 

We reconstruct  \Dz\ candidates in the \KPi\ and \KPiPiPi\
decay modes by combining charged tracks,
retaining combinations with an invariant mass within $2\sigma$
of the nominal \Dz\ mass $m_{\Dz}$.
In the $\Dz\ra\KPiPiz$ selection, the \piz\ candidates are
required to have a center-of-mass (CM) momentum  $p_{\piz}^*$
greater than $400$~\mevc.
For each $\KPiPiz$ combination, we use the kinematics of the decay
products and the known properties of the Dalitz plot for this
decay~\cite{dalitz} to compute the square of the decay amplitude
${\mathcal A}^2$.
We select combinations with ${\mathcal A}^2$ greater than $5\%$ of its
maximum value. This requirement selects mostly the $K^{-} \rho^{+}$ region
of the Dalitz plot. It rejects $62\%$ of the combinatorial
background, while keeping $76\%$ of $\Dz\ra\Km\pip\piz$ signal, as measured with the
$\Bp\ra\Dzb\pi^+$ sample.
Combinations with invariant mass within
25~\mevcc~($2.5\sigma$) of $m_{\Dz}$ are retained.

The \Dstarz\ candidates are selected from combinations of a \Dz\
and a \piz\ with $p_{\piz}^*>70$~\mevc. After kinematically
constraining \Dz\ and \piz\ candidates to their nominal masses, we
select the candidates with a mass difference $\Delta m \equiv
|m(\Dstarz)-m(\Dz)-142.2~\mevcc| <3.3$~\mevcc~($3\sigma$).

Two standard kinematic variables are used to select \Bz\
candidates: the energy-substituted mass
$\mes c^2\equiv\sqrt{(\frac{1}{2}s+c^2\mathbf{p}_\Upsilon\cdot
\mathbf{p}_{B})^2/E_\Upsilon^2-c^2\mathbf{p}_B^2}$ and the energy
difference $\DeltaE\equiv E_B^*-\frac{1}{2}\sqrt{s}$, where the
asterisk denotes the CM frame, $s$ is the square of the total
energy in the CM frame, $\mathbf{p}$ and $E$ are, respectively,
three-momentum and energy, and the subscripts $\Upsilon$ and $B$
refer to \Y4S\ and \Bz.
In calculating $\mathbf{p}_{B} $ and $E_B^*$ we constrain the mass of
the \DDstarz\ and \KS\ candidates to their 
respective nominal values.  
For signal events, $\mes$ is centered around the $\Bz$ mass with
a resolution of about 2.6~\mevcc, dominated by 
knowledge of the $e^+$ and $e^-$  beam energies. 
In simulated events the  \DeltaE\
resolution is found to be $\approx13$~\mev\ for all \Bz\ decay
modes considered in this analysis. The \Bz\ candidates are required to have $\mes > 5.2\gevcc$
and $|\DeltaE|< 100~\mev$.

We use two variables to reject most of the remaining background,
which is dominated by continuum events:
a Fisher discriminant~\cite{fisher} based on the energy flow in the
event and the polar angle
$\theta^*_B$ of the \Bz\ candidate in the CM frame.
For correctly reconstructed $B$ candidates $\cos{\theta^*_B}$ follows a
$1-\cos^{2}{\theta^*_B}$ distribution, whereas it is uniformly distributed
for continuum events and combinatorial background.
We require $|\cos\theta^*_B|<0.75$ for $\Bzb\ra\Dzb\Kstarzb$, and
$|\cos\theta^*_B|<0.85$ for all other decay modes. 
The Fisher discriminant \fisher\ is defined as a linear
combination of $|\cos{\theta_{TB}^*}|$ and two energy-flow moments
${\mathcal L}^0$ and ${\mathcal L}^2$. The variable
$\theta_{TB}^*$ is the angle in the CM frame between the thrust
axis~\cite{thrust} of the decay products of the \Bz and the thrust axis of all charged and neutral
particles in the event excluding the ones that form the \Bz. The
energy-flow moments ${\mathcal L}^0$ and ${\mathcal L}^2$ are
defined as ${\mathcal L}^i\equiv\sum_{j} p_{j}^*\
\cos^{i}{\theta_j}$ where $p^*_j$ is the CM momentum and
$\theta_j$ is the angle between the direction of particle $j$ with
respect to the thrust axis of the \Bz\ candidate, and the sum is
over all particles in the event (excluding those that form the
\Bz).
The requirement on \fisher\ varies for each decay channel because
of different levels of expected background. In the $\DDstarz\KS$ and
$\Dz\Kstarzb$ final states our requirement 
has an efficiency of about 80\% for the signal
while rejecting approximately 85\% of the
background; in the $\Bzb\ra\Dzb\Kstarzb$
mode a tighter requirement rejects 95\% of the background and has a signal efficiency of
55\%.

In the $\Dstarz\KS$ final state, approximately 5\% of the events
that satisfy all selection criteria contain more than one \Bz\
candidate. We retain the candidate with the smallest $\chi^2$
computed from the measured value of $m(\Dz)$ and
$m(\Dstarz)-m(\Dz)$, their nominal values, and their resolutions
in data. In the $\Dz\KS$, $\Dz\Kstarzb$, and $\Dzb\Kstarzb$ final
states we retain all selected \Bz\ candidates since the fraction of events 
with two or more candidates is negligible~($<1\%$).

The selected $\Bzb\ra \DDstarz \KKstarzb$ candidates include small
contributions from numbers of $B$ decays to similar final states which are
misreconstructed as signal candidates. We have studied these
backgrounds with large samples of simulated events, corresponding
to between 100 and 1000 times the size of our data sample,  for
the following categories of decays: 
$(1)$
$\Bzb\ra\Dz\rho^0,\rho^0\ra\pip\pim$ decays, where one of the two
pions is misidentified as a charged kaon; 
$(2)$ 
$\Bzb\ra\Dp\pim$ decays followed by  Cabibbo-suppressed decays
$\Dp\ra\KKstarzb\Kp$, and $\Bzb\ra\Dp\Km$ followed by $\Dp\ra\KKstarzb\pip$,
reconstructed in the $\Dz(\Km\pip)\KKstarz$ final states; 
$(3)$
charmless $\Bzb\ra\Km\pip\KS(n\pi)$ where the \Km\ and \pip\ are
wrongly combined to form a $\Dz\ra\Km\pip$ candidate; 
$(4)$
$\Bzb\ra\Dstarzb\KKstarz$, $\Dstarzb\ra\Dzb\gamma$ candidates, where
a low energy photon is not reconstructed; 
$(5)$ 
the decays $\Bm\ra\Dstarz\Km, \Dstarz\ra\Dz\piz/\gamma$, 
$\Bm\ra\Dz\Kstarm,\Kstarm\ra\Km\piz,\KS\pim$, and 
$\Bz\ra\Dstarm\Kp, \Dstarm\ra\Dzb\pim$, where a low-energy \piz, \pim, 
or photon is replaced by a random low-momentum charged particle.
The contribution of category $(1)$ is found to be less than 0.01
events and  hence is neglected.  The contribution of category $(2)$
is also negligible in all modes, except for $B\ra\Dz\Kzb,
\Dz\ra\Km\pip$. We eliminate 87\% of these background events by requiring the
invariant masses $m(\KS\Kp)$ and $m(\KS\pip)$ to be more than
20~\mevcc\ away from the nominal \Dp\ mass.
The \mes\ spectrum of the remaining background events in this
category, and in categories $(3)$--$(5)$, show a broad enhancement
near the $B$ mass.
However, due to the \Dz\
mass constraint,  \Bz\ candidates with a misreconstructed \Dz\ do
not peak, unlike the signal, in the \DeltaE\ distribution at zero.
In the decay $\Bzb\ra\Dzb\Kstarzb$, the charge correlation used in
the selection removes all contributions from known $B$ decays
included in simulated events.

The signal yield for each \Bz\ decay mode is determined with a
two-dimensional extended unbinned maximum likelihood fit to the \mes\ and
\DeltaE\ distributions, separately for each \Dz\ decay mode.
The probability density function~(PDF) is a sum of three components:
a signal component ${\mathcal G}(\mes)\times{\mathcal G}(\DeltaE)$, a
background component ${\mathcal G}(\mes)\times{\mathcal P}_{1}(\DeltaE)$, 
accounting for other $B$ decays misreconstructed as signal, and
a combinatorial background component ${\mathcal T}(\mes)\times{\mathcal P}_{2}(\DeltaE)$.
Here, ${\mathcal G}(\mes)$ is a Gaussian describing the \mes\ distribution of signal
and misreconstructed $B$ decays;  ${\mathcal G}(\DeltaE)$  is a Gaussian describing
the signal \DeltaE\ distribution; 
${\mathcal P}_{i}(\DeltaE)$ are first-order polynomials describing the \DeltaE\
distributions of background events.
The \mes\ distribution of the combinatorial background is 
parameterized by a  threshold function ${\mathcal T}(\mes)$ 
defined as ${\mathcal T}(\mes)\sim
\mes \sqrt{1-x^2}\exp\{-\xi(1-x^2)\}$~\cite{argus}, where
$x=2\mes c^2/\sqrt{s}$ and $\xi$ is a shape parameter.
The mean and the resolution of ${\mathcal G}(\mes)$ and ${\mathcal G}(\DeltaE)$ 
are fixed to values measured in the $\Bp\ra\Dzb\pip$ calibration sample.

\begin{table}[!thb]
\caption{ Signal yield $N_S$, signal significance ${\mathcal S}$, 
effective signal efficiency $\varepsilon_{\rm eff}$, 
and the measured branching fraction ${\mathcal B}$ for 
the $\Bztilde\ra\DDstarz\Kztilde$,  $\Bzb\ra\Dz\Kstarzb$, and $\Bzb\ra\Dzb\Kstarzb$ decays. 
The efficiency $\varepsilon_{\rm eff}$ is defined as
$\sum_i \varepsilon_i\times{\mathcal B}_i$, where the sum is over the \Dz\ decay modes, 
$\varepsilon_i$ are the signal
reconstruction efficiencies, and ${\mathcal B}_i$ are the corresponding
intermediate  branching fractions  for \Dstarz, \Dz, \Kstarz, and
\Kz\ decays to final states reconstructed in this analysis.}
\begin{center}
\begin{tabular}{lcccc}
\hline
\hline
&\\[-9pt]
 $B$ Mode & $N_{S}$  & ${\mathcal S}$ &
   $\varepsilon_{\rm eff}~[\%]$ & ${\mathcal B}~[10^{-5}]$  \\[1pt]
\hline
&\\[-9pt]
$\Bztilde\ra\Dz\Kztilde$        &  $104\pm14   $ & $9.2\sigma$ &
$0.82$ & $\dzkzBrVal\pm\dzkzBrStat\pm\dzkzBrSyst$\\
$\Bztilde\ra\Dstarz\Kztilde$    &  $17.1\pm5.2$ & $4.3\sigma$ & $0.17$ &
$\dstarzkzBrVal \pm \dstarzkzBrStat\pm \dstarzkzBrSyst$\\
$\Bzb\ra\Dz\Kstarzb$ &  $77\pm12$ & $7.9\sigma$ &  $0.84$ &
$\dzbkstarzBrVal \pm \dzbkstarzBrStat\pm \dzbkstarzBrSyst$\\
$\Bzb\ra\Dzb\Kstarzb$&  $-3.6^{+6.8}_{-5.5}$ & -- & $0.47$ &
$\dzkstarzBrVal \pm \dzkstarzBrStat\pm \dzkstarzBrSyst$\\
&\\[-9pt]
\hline
\hline
\end{tabular}

\label{tab:yields}
\end{center}
\end{table}

\begin{figure}[!htb]
\begin{center}
\epsfig{file=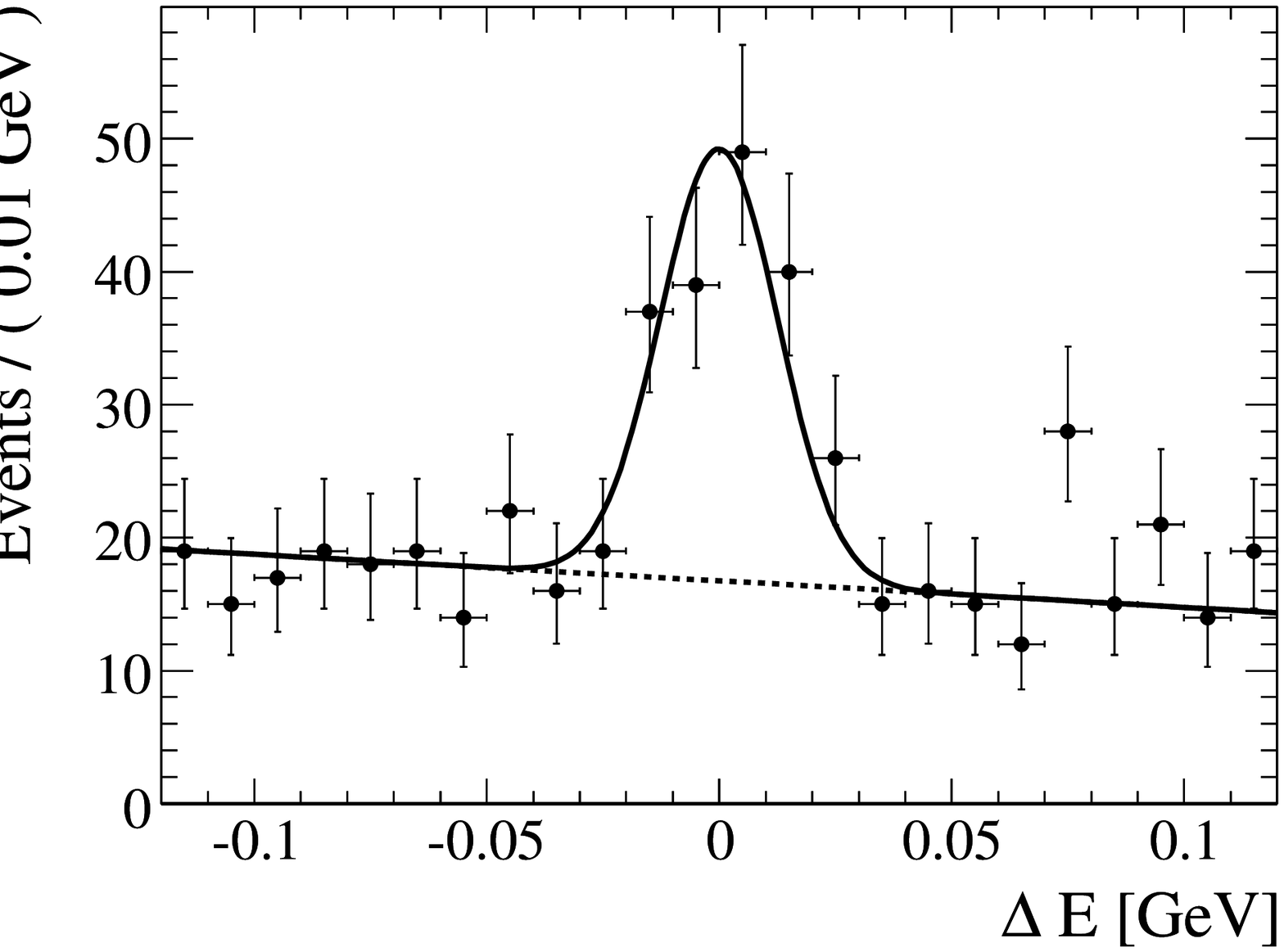,width=0.49\linewidth}
\put(-102,74){a)}
\epsfig{file=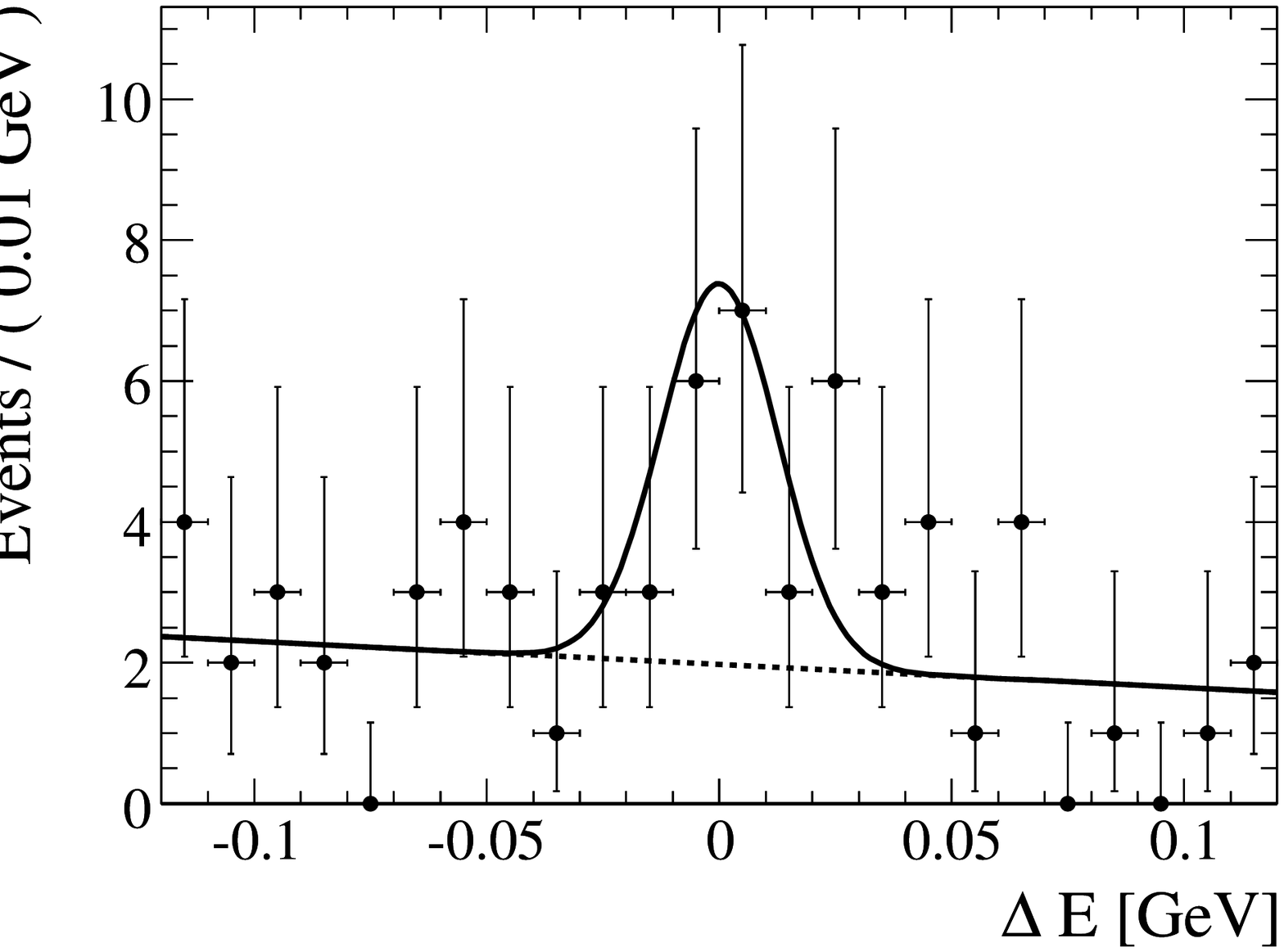,width=0.49\linewidth}
\put(-102,74){b)}\\
\epsfig{file=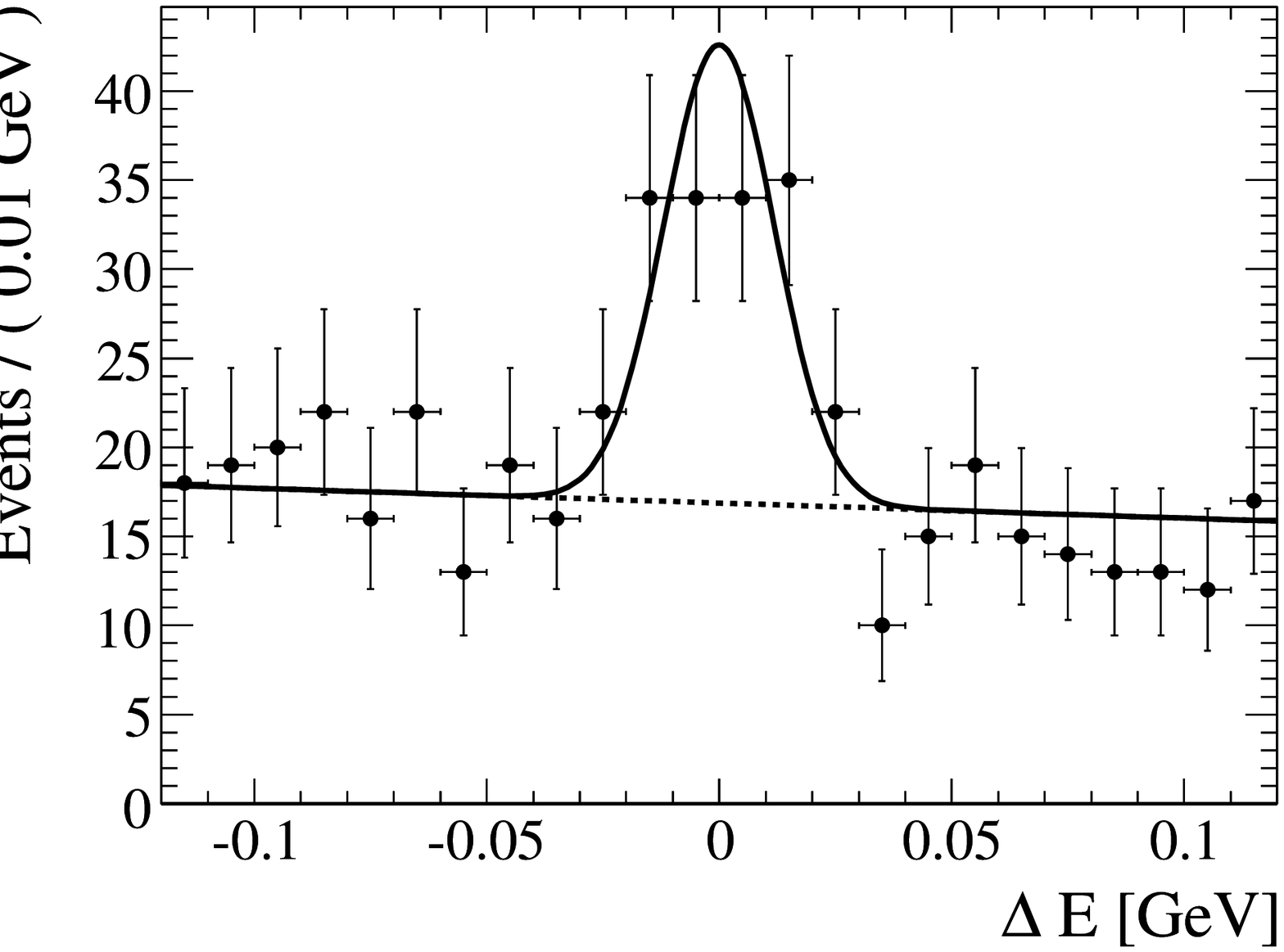,width=0.49\linewidth}
\put(-102,74){c)}
\epsfig{file=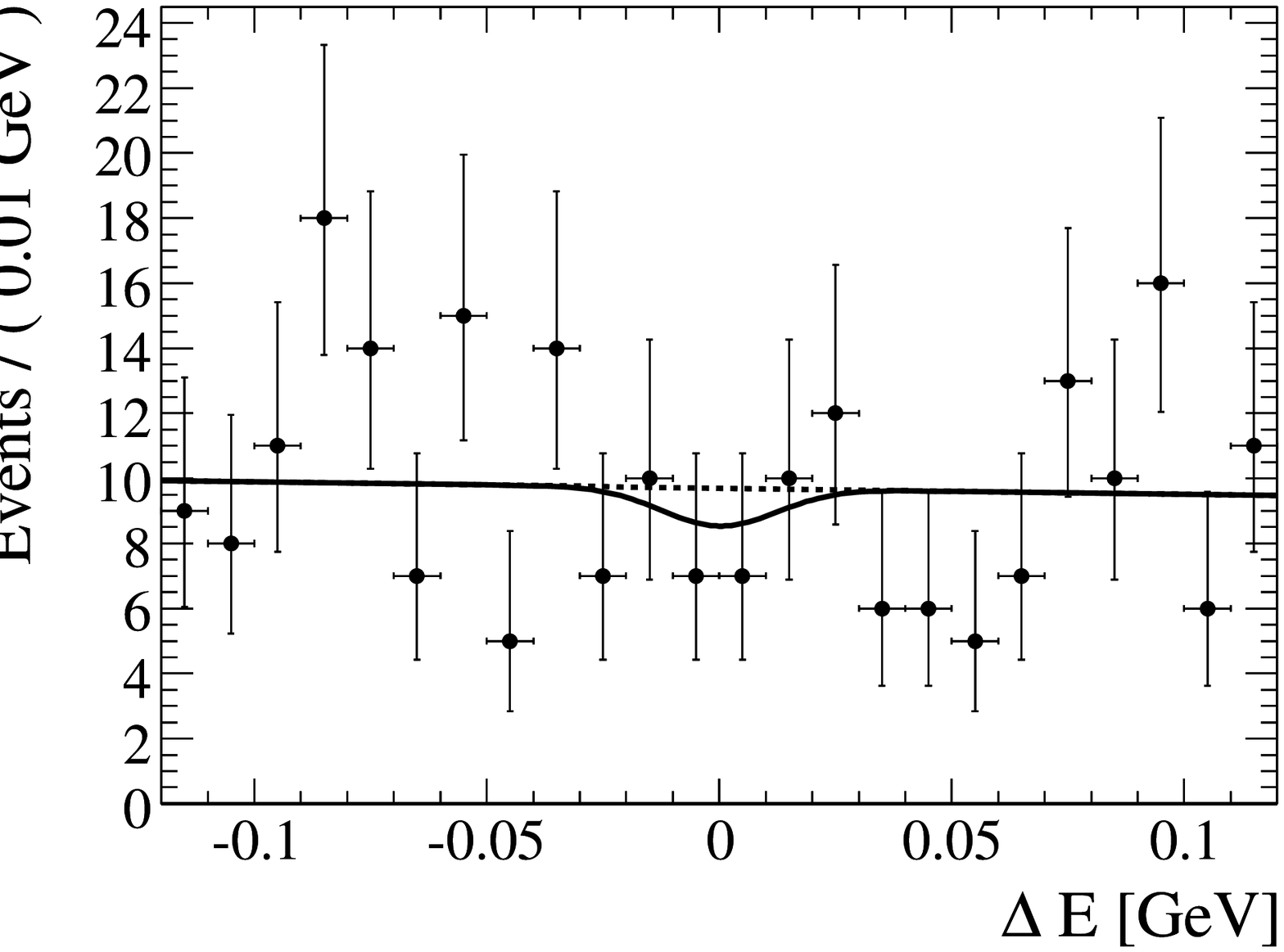,width=0.49\linewidth}
\put(-102,74){d)}
\caption{
Distribution of \DeltaE\ for  a) $\Bztilde\ra\Dz\Kztilde$, b)
$\Bztilde\ra\Dstarz\Kztilde$, c) $\Bzb\ra\Dz\Kstarzb$, and d)
$\Bzb\ra\Dzb\Kstarzb$ candidates with $|\mes-5280~\mevcc|< 8\mevcc$. The
points are the data, 
the solid curve is the projection of the likelihood fit, and
the dashed curve represents the background component.
}
\label{fig:yields}
\end{center}
\end{figure}
The measured signal yields are summarized in
Table~\ref{tab:yields}. The \DeltaE\ distributions of 
candidates with $|\mes-5280|<8$~\mevcc for the
sums of the reconstructed \Dz\ decay modes are illustrated in
Figure~\ref{fig:yields}.
The signal significance ${\mathcal S}$  is computed as
${\mathcal S} = \sqrt{2(\ln{\mathcal L}(N_{S})-\ln{\mathcal L}(N_{S}=0))}$, where
${\mathcal L}(N_{S})$ is the maximum likelihood of the nominal fit, and 
${\mathcal L}(N_{S}=0)$ is the value obtained after repeating the fit with the
signal yield $N_{S}$ constrained to be zero.

The branching fraction ${\mathcal B}$ for each \Bz\ decay mode is
the weighted average of the branching fractions
${\mathcal B}_{j}$ in each \Dz\ channel $D_j=\{\KPi,\KPiPiPi,\KPiPiz\}$,
computed as
\begin{eqnarray}
{\mathcal B}_{j}  = \frac{ N_{S_{j} }}
                       { 2 \times
                         N_{\BB} \times
                         {\mathcal B}(\Y4S\ra\Bz\Bzb) \times
                         {\mathcal B}_{D_j} \times
                         {\mathcal B}_K \times
                         \varepsilon_{j}
                       }
\label{eq:br expr}
\end{eqnarray}
where $N_{S_{j}}$ is the signal yield from the likelihood fit,
$N_{\BB}$ is the total number of $\Y4S\ra \BB$ events, ${\mathcal
B}_{D_j}$ is the branching fraction ${\mathcal B}(\Dz\ra D_j)$ in
$\Bztilde\ra\Dz\KKstarztilde$ and ${\mathcal
B}(\Dstarz\ra\Dz\piz)\times{\mathcal B}(\Dz\ra D_j)$ in
$\Bztilde\ra\Dstarz\Kztilde$, ${\mathcal B}_K $ is the
$\Kz\ra\KS\ra\pip\pim(\Kstarz\ra\Kp\pim)$ branching fraction in
$\Bztilde\ra\DDstarz\Kztilde(\Bzb\ra\Dz\Kstarzb, \Dzb\Kstarzb)$, and $\varepsilon_{j}$ is the signal
reconstruction efficiency.
We assume ${\mathcal B}(\Y4S\ra\Bz\Bzb)=0.5$.
The systematic uncertainties for the branching fractions
include contributions  from estimated misreconstructed $B$
background~(1--13\%)~\cite{bback}, 
variation of parameters kept fixed in the likelihood fit~(2--8\%), 
\DDstarz\ branching
fraction~(2.4--6.9\%), \piz\ reconstruction efficiency~(3\%),
photon reconstruction efficiency~(1.8\%), 
charged-track reconstruction efficiency~(0.8\% per
track), simulation statistics~(1--4\%), efficiency correction
factors~(1--4\%), kaon identification efficiency~(2\% per kaon), \KS\
reconstruction efficiency~(1.6\%), and the number of
$\BB$ events~(1.1\%).
The efficiency correction factors are obtained by  comparing  data with MC simulation
in the $\Bp\ra\Dzb\pip$ control sample.
The largest contributions to the uncertainties
in these factors are from selection requirements for the 
\piz\ momentum $p^*_{\piz}$ and the amplitude $|{\mathcal A}|^2$ 
in the $\Dz\ra\Km\pip\piz$ decay and the 
Fisher discriminant \fisher.
We measure 
\begin{eqnarray}
\BR(\Bztilde\ra\Dz\Kztilde) &=& (\dzkzBrVal\pm\dzkzBrStat
\pm\dzkzBrSyst)\brscale \nonumber \\
\BR(\Bztilde\ra\Dstarz\Kztilde) &=& (\dstarzkzBrVal \pm \dstarzkzBrStat
\pm \dstarzkzBrSyst)\brscale \nonumber  \\
\BR(\Bzb\ra\Dz\Kstarzb) &=& (\dzbkstarzBrVal \pm \dzbkstarzBrStat
\pm\dzbkstarzBrSyst)\brscale \nonumber \\
\BR(\Bzb\ra\Dzb\Kstarzb) &=& (\dzkstarzBrVal \pm \dzkstarzBrStat
\pm \dzkstarzBrSyst)\brscale \nonumber
\end{eqnarray}
where the uncertainties are, respectively, statistical and systematic.
For the decay  $\Bzb\ra\Dzb\Kstarzb$ we use the Bayesian method to
compute the upper limit $N_{UL}$ on the observed number of events.
The value of $N_{UL}$ at 90\% C.L. is defined as 
$\int^{N_{UL}}_0  {\mathcal L}(N)\
dN=0.9$, where ${\mathcal L}(N)$ is the maximum  likelihood
function from the fit to the \mes\ and \DeltaE\ distributions. We
assume a flat prior probability density function for ${\mathcal
B}>0$.
We account for systematic uncertainties by numerically convolving
${\mathcal L}(N)$ with a Gaussian distribution with a width determined
by the relative systematic uncertainty multiplied by the measured
signal yield.
We obtain $\BR(\Bzb\ra\Dzb\Kstarzb) < \dzkstarzLim\times 10^{-5}$ at 90\% C.L.

We compute an upper limit on the ratio $\rbk$  by measuring the ratio ${\mathcal R}_i$ 
in each \Dz\ decay mode. We use the expression 
${\mathcal R}_i = (\varepsilon_{D_i\Kbar}/\varepsilon_{\Dbar_i\Kbar}) \cdot 
(N_{\Dbar_i\Kbar}/N_{D_i\Kbar})$ to obtain the PDF
 for ${\mathcal R}_i$ from the unbinned maximum likelihood fit 
described earlier.
In this expression 
$\varepsilon_{\Dbar_i\Kbar}$~($\varepsilon_{D_i\Kbar}$) and
$N_{\Dbar_i\Kbar}$~($N_{D_i\Kbar}$) are, respectively, the reconstruction efficiency and
fitted yield of the $\Bzb\ra\Dzb\Kstarzb, \Dzb\ra\Km X^+_i$~($\Bzb\ra\Dz\Kstarzb, \Dz\ra\Km X^+_i$)
decay modes.
The uncertainties on $\varepsilon_{\Dbar_i\Kbar}$, $\varepsilon_{D_i\Kbar}$, and 
$N_{D_i\Kbar}$ are used to obtain the
posterior PDF ${\mathcal L}({\mathcal R}_i)$ for each ${\mathcal R}_i$.
We assume a Gaussian PDF for $r_{D_i}$. We compute the PDF for \rbk\ by convolving 
${\mathcal L}({\mathcal R}_i)$ and $r_{D_i}$ according to equation ~(\ref{R_ADS}).
We obtain the limit 
$\rbk < 0.40$ at 90\% C.L.\
with a Bayesian method using uniform priors for ${\mathcal R}_i>0$ 
and by taking into account the full range
$0^{\circ}$--$180^{\circ}$ for $\gamma$ and $\delta_i$.
The present signal yields combined with this limit on \rbk\ suggest that 
a substantially larger data sample is needed for a competitive
time-dependent measurement of \stwobg\ in $\Bztilde\ra\DDstarz\Kztilde$ decays.

In summary, we have presented 
measurements of the branching fractions for the decays  $\Bztilde\ra\Dz\Kztilde$ and 
$\Bzb\ra\Dz\Kstarzb$,
evidence for the decay $\Bztilde\ra\Dstarz\Kztilde$, and an upper limit for the ratio 
\rbk. Our results are in agreement
with previous measurements of these modes~\cite{belle-d0ks-hepex}.

We are grateful for the excellent luminosity and machine conditions
provided by our \pep2\ colleagues, 
and for the substantial dedicated effort from
the computing organizations that support \babar.
The collaborating institutions wish to thank 
SLAC for its support and kind hospitality. 
This work is supported by
DOE
and NSF (USA),
NSERC (Canada),
IHEP (China),
CEA and
CNRS-IN2P3
(France),
BMBF and DFG
(Germany),
INFN (Italy),
FOM (The Netherlands),
NFR (Norway),
MIST (Russia), and
PPARC (United Kingdom). 
Individuals have received support from CONACyT (Mexico), 
Marie Curie EIF (European Union),
the A.~P.~Sloan Foundation, 
the Research Corporation,
and the Alexander von Humboldt Foundation.


\begin{thebibliography}{99}
\bibitem{charge-conj}
Charge conjugation is implied throughout this paper, unless
explicitly stated otherwise.
\bibitem{bback}
The contribution of misreconstructed $B$ background in the
$\Bzb\ra\Dzb\Kstarzb$ mode, where no signal is observed, is about two
events.
\end{thebibliography}
\end{document}